\documentclass[10pt]{article}
\usepackage{graphicx}

\def\Title#1{\begin{center} {\Large #1 } \end{center}}
\def\Author#1{\begin{center}{ \sc #1} \end{center}}
\def\Address#1{\begin{center}{ \it #1} \end{center}}

\newenvironment{Abstract}{\begin{quotation} \begin{center} 
             \large ABSTRACT \end{center}\bigskip 
      \begin{center}\begin{large}}{\end{large}\end{center} \end{quotation}}

\newenvironment{Presented}{\begin{quotation} \begin{center} 
             PRESENTED AT\end{center}\bigskip 
      \begin{center}\begin{large}}{\end{large}\end{center} \end{quotation}}

\def\beq{\begin{equation}}
\def\eeq#1{\label{#1}\end{equation}}
\def\eeqn{\end{equation}}

\def\beqa{\begin{eqnarray}}
\def\eeqa#1{\label{#1}\end{eqnarray}}
\def\eeqan{\end{eqnarray}}

\let\bar=\overbar

\def\Dslash{\not{\hbox{\kern-4pt $D$}}}
\def\dslash{\not{\hbox{\kern-2pt $\del$}}}

\def\msb{{\bar{\ssstyle M \kern -1pt S}}}

\usepackage{color}
\usepackage{amsmath}
\usepackage{paralist}

\def\affiliation{
for the ALICE Collaboration, \\
CERN \\
Geneva, Switzerland}

\begin{document}

% large size for the first page
\large
\begin{titlepage}
%\pubblock

%% Change the title, name, abstract
%% Title 
\vfill
\Title{  Tracking performance in high multiplicities environment at ALICE  }
\vfill

%  if you need to add the support use this, fill the \support definition above. 
%   \Author{ FIRSTNAME LASTNAME \support }
\Author{ David Rohr }
\Address{\affiliation}
\vfill
\begin{Abstract}
In LHC Run 3, ALICE will increase the data taking rate significantly to 50\,kHz continuous read out of minimum bias Pb-Pb events.
This challenges the online and offline computing infrastructure, requiring to process 50 times as many events per second as in Run 2, and increasing the data compression ratio from 5 to 20.
Such high data compression is impossible by lossless ZIP-like algorithms, but it must use results from online reconstruction, which in turn requires online calibration.
These important online processing steps are the most computing-intense ones, and will use GPUs as hardware accelerators.
The new online features are already under test during Run 2 in the High Level Trigger (HLT) online processing farm.
The TPC (Time Projection Chamber) tracking algorithm for Run 3 is derived from the current HLT online tracking and is based on the Cellular Automaton and Kalman Filter.
HLT has deployed online calibration for the TPC drift time, which needs to be extended to space charge distortions calibration.
This requires online reconstruction for additional detectors like TRD (Transition Radiation Detector) and TOF (Time Of Flight).
We present prototypes of these developments, in particular a data compression algorithm that achieves a compression factor of~9 on Run 2 TPC data, and the efficiency of online TRD tracking.
We give an outlook to the challenges of TPC tracking with continuous read out.
\end{Abstract}
\vfill

% DO NOT CHANGE 
\begin{Presented}
The Fifth Annual Conference\\
 on Large Hadron Collider Physics \\
Shanghai Jiao Tong University, Shanghai, China\\ 
May 15-20, 2017
\end{Presented}
\vfill
\end{titlepage}
\def\thefootnote{\fnsymbol{footnote}}
\setcounter{footnote}{0}
%

% normal size for the rest
\normalsize 

%% Your paper should be entered below. 

\section{Introduction}

The ALICE experiment~\cite{alice} is the dedicated large-scale heavy ion experiment at the LHC at CERN.
During the heavy ion phases at the end of the year, ALICE has taken Pb-Pb collisions with up to 500\,Hz trigger rate during Run 1 and up to~1\,kHz in Run 2, and p-Pb interactions with up to~2\,kHz.
ALICE has demonstrated good tracking performance in the barrel region~\cite{aliceperf} in terms of matching efficiency between the detectors, track and vertex resolution, etc., which has remained the same during Run 1 and Run 2.
For Run 3 after the Long Shutdown 2, ALICE will receive a major upgrade, which will enable minimum bias Pb-Pb data taking with all events read out at~50\,kHz interaction rate, the design rate of the LHC after the upgrade.
This involves major upgrades of several detectors, in particular the TPC, which is the main tracking detector of ALICE.
The wire chambers, which limit the triggered read-out rate to few kHz, will be replaced by gas electron multipliers (GEM) enabling continuous read out~\cite{tpctdr}.

\looseness=-1
In order to cope with the data rate expected during Run 3, the entire online and offline computing infrastructure of ALICE will be upgraded, which includes a new large scale online computing farm~\cite{o2tdr}.
The constraint for the processing is to maintain the tracking efficiency and resolution achieved during Run 1 and 2 under Run 3 conditions.
ALICE will take continuous minimum bias data without triggers and store all events.
Thus, compared to Run 2, about 50 times as many Pb-Pb events must be reconstructed and stored.
Even with the most optimistic estimations for storage costs during Run 3, it is prohibitively expensive to store one order of magnitude more data as today and perform traditional offline analysis afterwards.
The online processing must perform elaborate data compression and the TDR foresees a compression factor of~$\approx 20$ for TPC raw data.
The final size is fixed, so the exact factor may change with the raw data format.
Such compression is infeasible using lossless ZIP-like compression, but the compression must rely on results of the reconstruction.
Moreover, sufficiently precise real-time reconstruction of the drift detectors  that are sensitive to environmental conditions requires online calibration.
By design, the online reconstruction must process 50 times more events than today in the same time.
The multiplicity in the TPC will be much higher, with in average five Pb-Pb events per drift time of~$100$\,$\mu$s.
The increase in multiplicity and the absence of the gating grid in continuous readout mode also increase the space charge in the TPC and thus the distortions during the drift.

These constraints set the stage for the online--offline computing upgrade~\cite{o2tdr}: perform online calibration, reconstruction, and data compression with a compression factor 20 for 50 times as many events as today in continuous read out data with larger TPC distortions than today.
Since the current offline processing is too slow to achieve these tasks with the computing budget available for Run 3, track reconstruction will be derived from the online tracking in the High Level Trigger (HLT).
As prototype for Run 3, we attempt to implement many of the new features already in the current online HLT computing farm.
The remainder of this paper is structured as follows:
we will briefly introduce the current online tracking algorithm that we want to extend for Run 3.
Next, we will present our approaches to achieve the required compression factor and to perform the calibration online.
In the last part, we discuss the challenges for tracking during Run 3 and the required improvements for the online tracking.

\section{Online Tracking in ALICE}

\begin{figure}[t!]
\centering
\includegraphics[width=\textwidth]{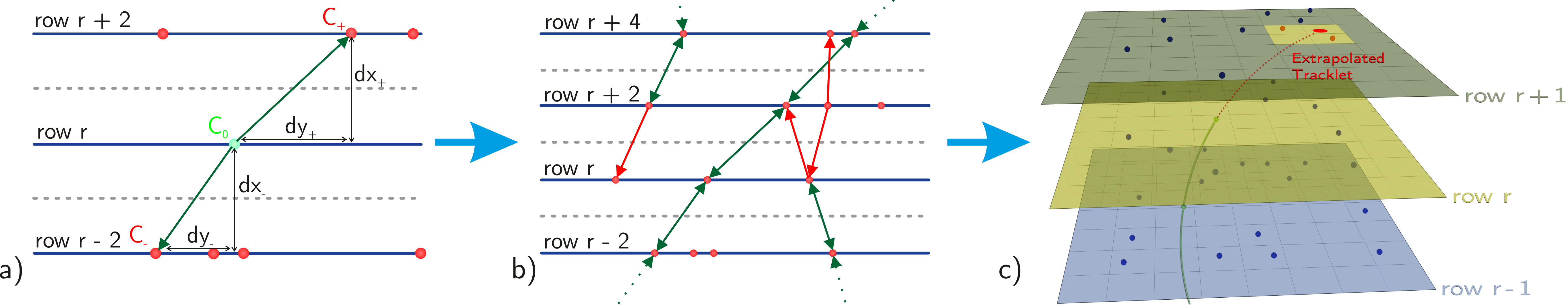}
\caption{ Illustration of Cellular Automaton seeding (a and b) and track following (c) of HLT tracking }
\label{fig:tracking}
\end{figure}

The ALICE HLT online tracking~\cite{tns} is based on the Cellular Automaton principle and the Kalman filter.
The TPC cylinder volume is divided into 36 sectors, and tracking has two phases.
The first phase is the sector tracker which finds track segments in all the TPC sectors individually.
It starts with a Cellular Automaton to find short track candidates of usually 4 to 10 TPC hits by forming (Fig.~\ref{fig:tracking}a) and concatenating (Fig.~\ref{fig:tracking}b) links of adjacent hits arranged close to a straight line.
The next step is the track following, which fits track parameters to the track candidate and extrapolates the trajectory through the TPC sector volume to pick up additional clusters and refine the fit using a simplified Kalman filter (Fig.~\ref{fig:tracking}c).
The last step of the sector tracker is the selection of good tracks and final assignment of clusters to tracks dissolving ambiguities.

\begin{figure}[b]
 \begin{minipage}[t]{0.495\textwidth}
  \centering
  \includegraphics[height=1.97in]{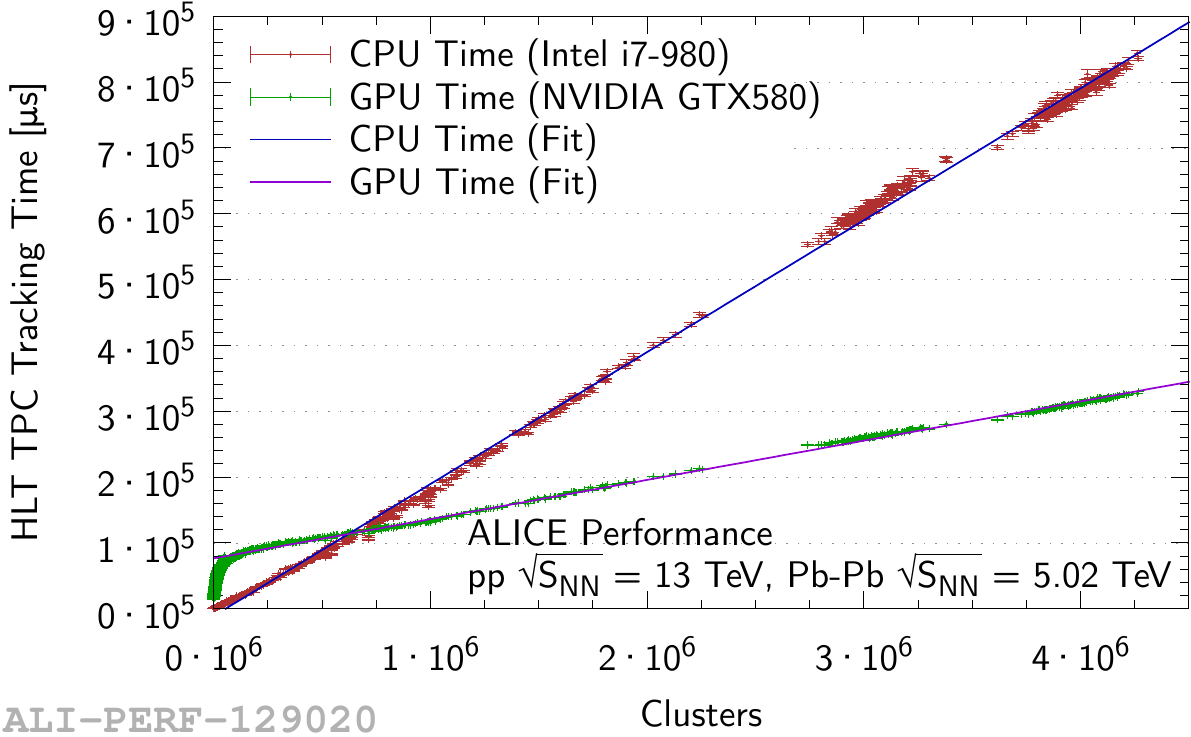}
  \caption{ Tracking time dependency on input size }
  \label{fig:tracking_linear}
 \end{minipage}
 \begin{minipage}[t]{0.495\textwidth}
  \centering
  \includegraphics[height=2in]{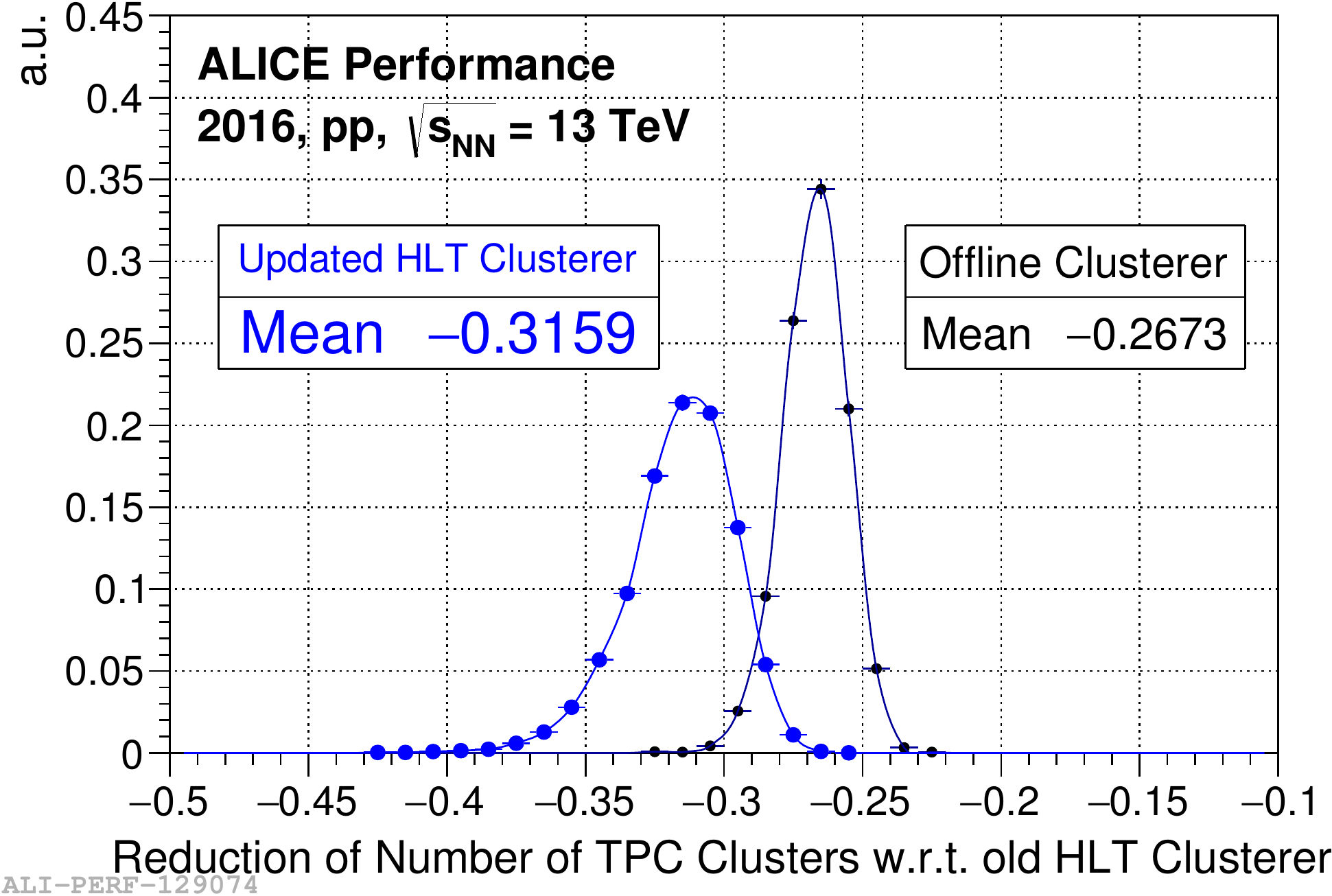}
  \caption{ Number of cluster reduction by new HWCF }
  \label{fig:hwcf_red}
 \end{minipage} 
\end{figure}

The second phase merges the track segments found in the individual sectors to final tracks and performs a track refit.
The first phase is computing-critical contributing more than~80\,\% to the computing time and was thus adapted to run on graphics cards (GPUs)~\cite{chep}.
GPU tracking results have been proven to be absolutely compatible to the CPU, and the GPU tracking is active in 24/7 operation since 2012.
The GPU tracker yields a speedup of about a factor 3 comparing CPUs and GPUs state of the art at the same time~\cite{cnna}.
Since the online computing farm is not dedicated to tracking but must perform other tasks as well, the GPUs increase the tracking throughput by a factor 10 compared to a CPU-only solution.
Perhaps even more important for increased multiplicity in Run 3 is the linear dependency on the input data size (see Fig.~\ref{fig:tracking_linear}).
This will enable us to run the tracking on large time-frames containing $\approx 20$\,ms of continuous read out data with 5 minimum bias Pb-Pb events per drift time.
An existing prototype for the final track fit on GPU is not yet in operation because the data transfer between host and GPU poses a bottleneck.
All GPU code is implemented in a generic way, such that the same code runs on the processor (possibly parallelized via OpenMP), on NVIDIA GPUs via CUDA, and on AMD GPUs via OpenCL~\cite{chep15}.
One HLT server needs~$110$\,ms per central Pb-Pb event, and the whole HLT farm can fully reconstruct~$40,000,000$ tracks per second using~$180$ AMD S9000 GPUs.
In order to load modern GPUs to the full extent, we are processing multiple events in parallel because even central Pb-Pb events are too small to offer enough parallelism on their own.

\begin{figure}[t!]
\centering
\includegraphics[height=1.7in]{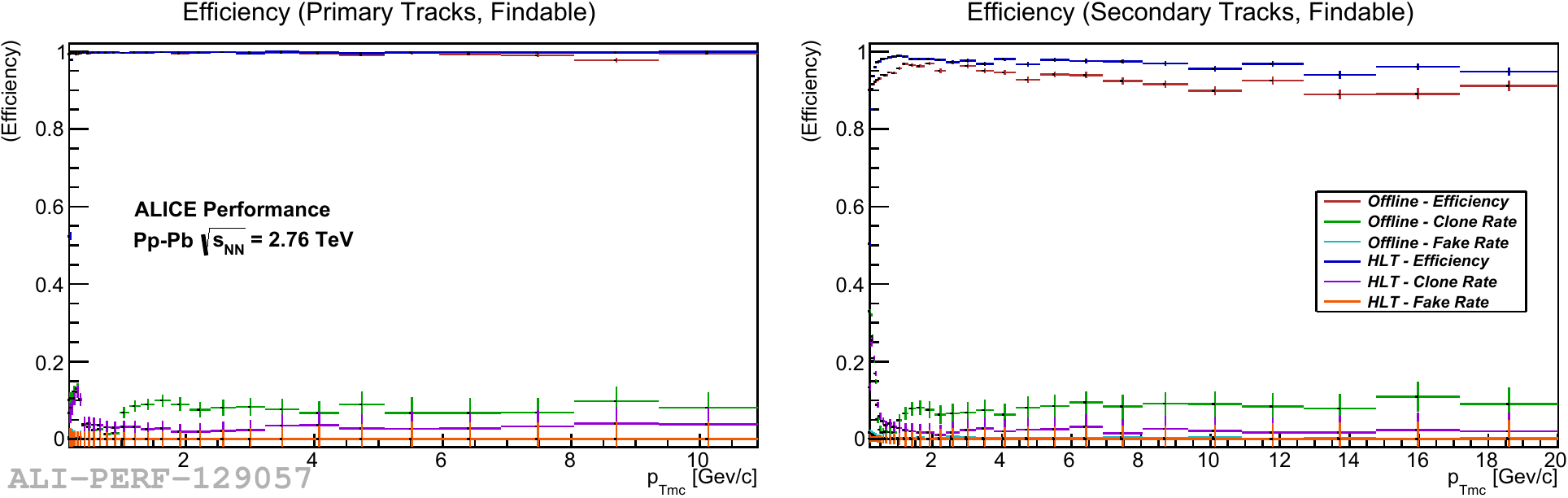} \\
\ \\
\includegraphics[height=3.4in]{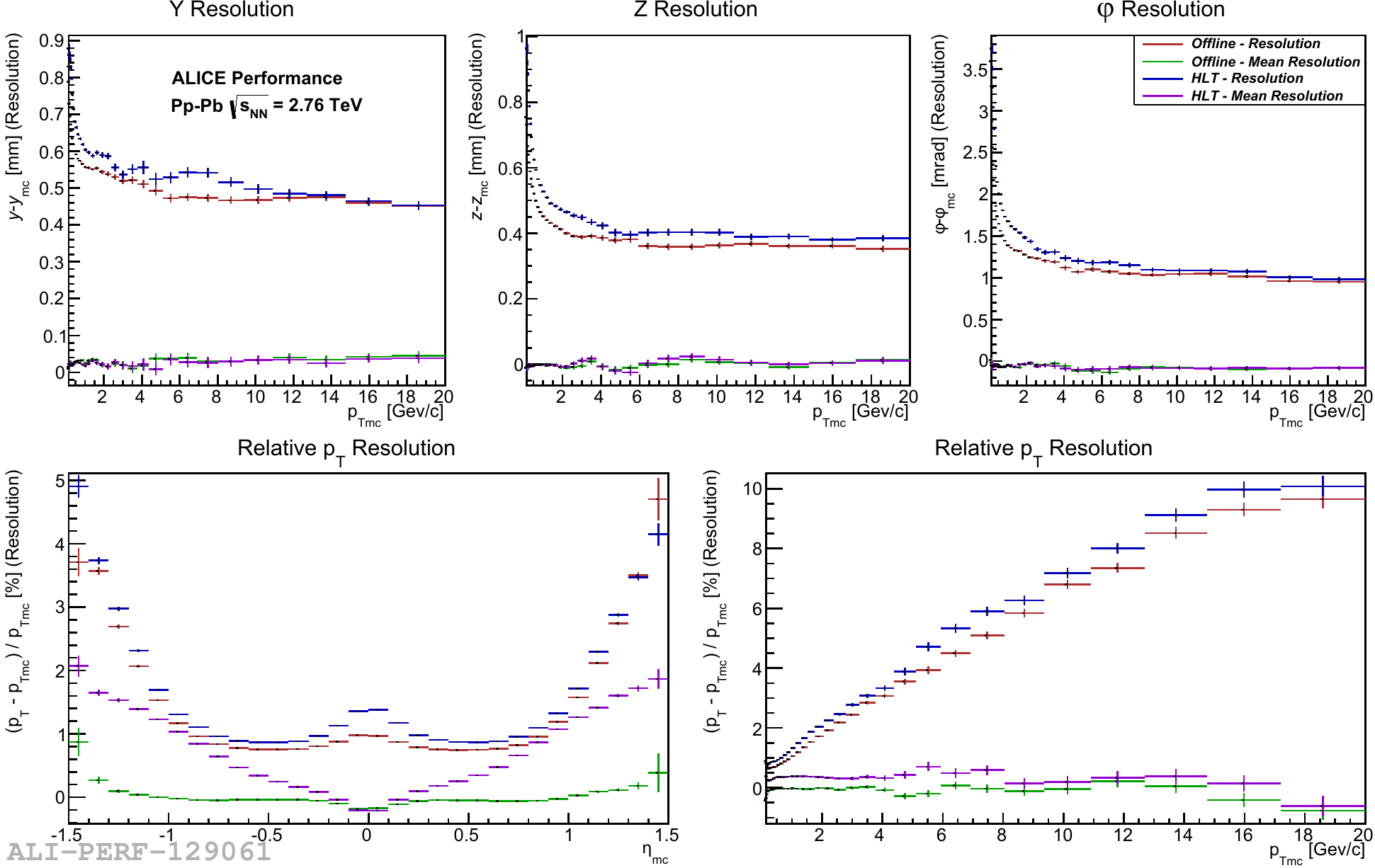}
\caption[Resolution and efficiency of ALICE online and offline tracking of TPC data only]{ Resolution and efficiency of ALICE online and offline tracking of TPC data only\\\textit{(Findable tracks range at least 70\,cm through the TPC, $P_{\text{T}} \geq 200$\,MeV, $\eta \leq 0.9$)}}
\label{fig:hlt_eff_res}
\end{figure}

\looseness=-1
In Fig~\ref{fig:hlt_eff_res} efficiency and resolution of online and offline tracker are compared on Monte-Carlo data.
The online tracking efficiency is absolutely compatible to offline.
The online version even features a lower clone rate due to the approach with segment finding and merging, and slightly better efficiency for secondaries since the Cellular Automaton seeding applies only a loose vertex constraint.
In terms of resolution, the offline version is always superior, and in particular for the transverse momentum resolution, the online version introduces a bias.
The reason is that online tracking does currently not respect the $B_x$ and $B_y$ components of the magnetic field leading to incorrect propagation depending on pseudorapidity (visible in the~$P_{\text{T}}$ resolution versus~$\eta$ plot).
These deficiencies in the resolution are investigated and will be addressed for Run 3.

\section{Data Compression}

\begin{figure}[t]
\centering
\includegraphics[height=1.8in]{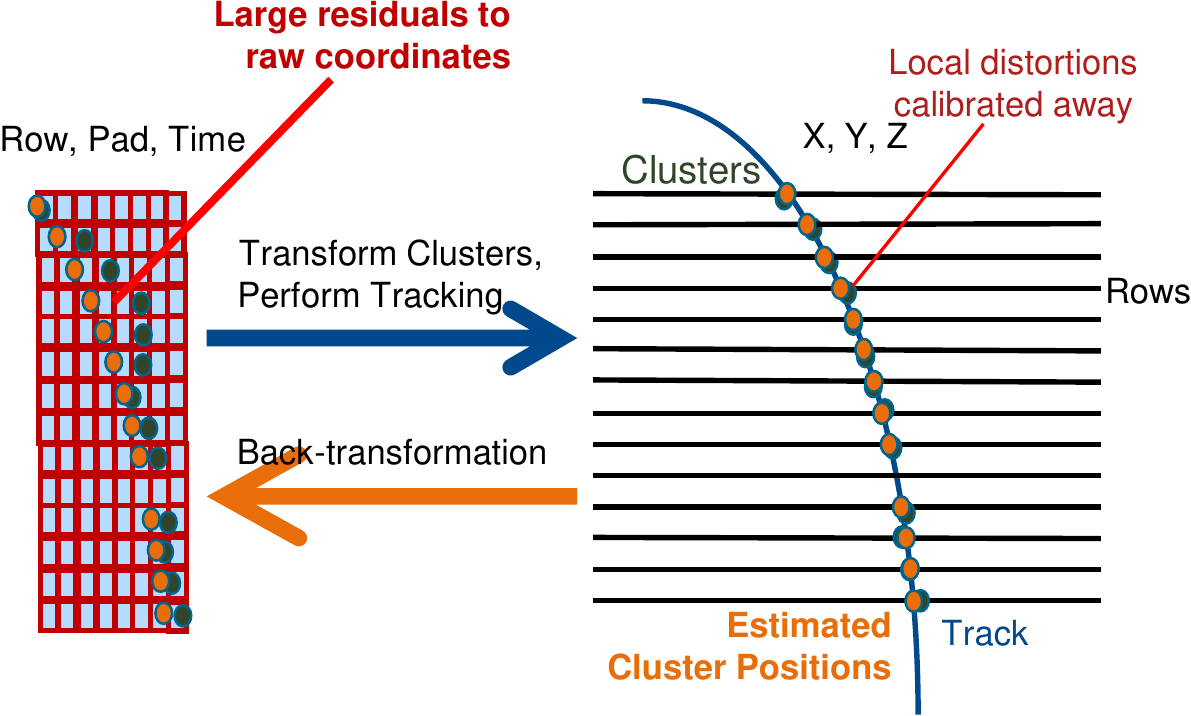}\ \ \  \includegraphics[height=1.8in]{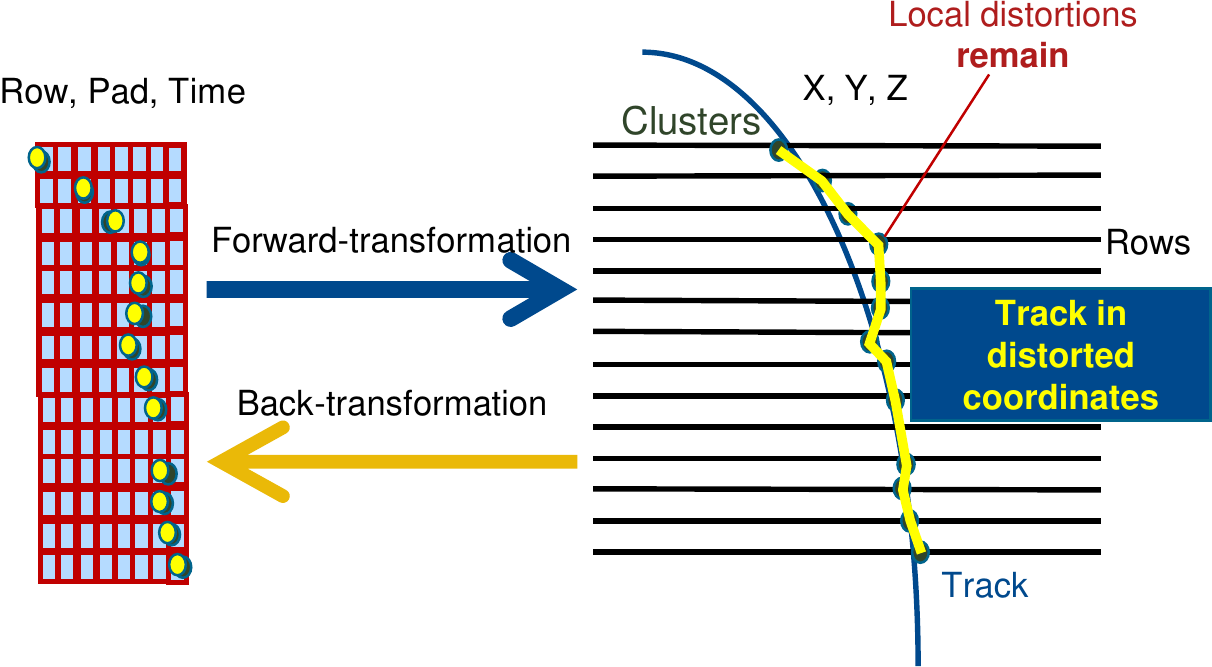}
\caption{ Illustration of track model compression. (left: old version, right: new version) }
\label{fig:trackmodel}
\end{figure}

We have started to gradually introduce more sophisticated compression algorithms to reach the required compression factor of 20.
The baseline is online hardware cluster finding (HWCF) in the FPGA-based read out card called C-RORC~\cite{crorc}, which provides the cluster properties to the online computing farm.
The cluster properties are then entropy-compressed using Huffman compression.
Beforehand, they are reformatted to have as little entropy as possible.
In the first iteration deployed in 2011, this reformatting converted them to a fixed-point integer format equaling the detector resolution achieving a total compression factor of around 4 in Run 1~\cite{run1compression}.
Since 2016 we are further reducing the entropy using an improved cluster format, which stores only position differences instead of absolute positions leading to smaller entropy for high occupancy.
This increases the compression ratio to~$5.5$ for pp and~$5.9$ for Pb-Pb~\cite{run2compression}.
We are currently developing a prototype for Run 3 compression, which we plan to deploy already during Run 2 in the HLT for verification, for which we evaluate the following steps on top:
\begin{compactitem}
\looseness=-1
 \item The HLT cluster finder is sensitive to noise which can lead to fake clusters if the noise surpasses the zero-suppression threshold.
 This is mostly pronounced in Run 2 data with argon gas in the TPC featuring a higher gain.
 We have implemented a noise filter and smoothing algorithm in the cluster finding, which reduces the number of clusters by~$32$\,\% on argon data compared to the old HLT cluster finder.
 Fig.~\ref{fig:hwcf_red} illustrates that the new version has even a better noise suppression than the offline cluster finder used as reference.
 All physics properties like track resolution, TPC-ITS (Inner Tracking System) matching, and d$E$/d$x$ performance of the new cluster finder were verified to be equally good as or better than with the old version.
 \item The fixed-point integer format is not ideal for compression of cluster-charge and width, where the relative precision is important instead of the absolute one.
 Thus, we truncate the least significant bits to zero reducing the precision to 4 non-zero bits for the charge and 3 bits for the width yielding a 15\,\% better compression.
 \item We have measured that arithmetic entropy encoding achieves in average 4.7\,\% better compression ratio than Huffman compression.
 \item Using information from the tracking, one can reduce the entropy further.
 For clusters attached to tracks, it is possible to store the residual of the cluster to the track instead of the absolute cluster position (or differences).
 These residuals are supposed to be small and have lower entropy.
 Unfortunately, the first implementation attempted already during Run 1 suffered from some deficiencies, yielding too large residuals, as explained in the following.
\end{compactitem}
Before the tracking, the clusters are transformed from TPC-native pad, row, and time coordinates to spacial ones.
The first version propagated the helix through the TPC volume and intersected it with the TPC read out rows to obtain the residuals to the TPC clusters in $x$, $y$, and $z$ coordinates.
Space charge distortions (SCD) are calibrated away as described in section~\ref{sec:scd_calib}.
In order to compress raw coordinates, these~$x$, $y$, and $z$ positions must be transformed back to pad, row, and time.
The accurate TPC transformation is not reversible, so the first approach used a linear model for the reverse transformation.
This is insufficient because the helix propagation and the back-transformation are not exact enough.
It ignores scattering and energy loss, and does not take into account SCD at all, which cause large residuals and will even increase in Run 3 (see upper illustration in Fig.~\ref{fig:trackmodel}).

Instead, we follow a more elaborate approach.
We approximate the TPC transformation with an invertible polynomial one.
This does not respect local effects like distortions and yields clusters in a distorted coordinate system.
We refit the track (using the cluster to track association from the tracking) in this distorted system, where the Kalman filter follows the local distortions.
After each extrapolation step to the next TPC row, we apply the reverse transformation for the extrapolated track position, and store the residual to the cluster in pad, row, and time coordinates (see Fig.~\ref{fig:trackmodel}).
For decompression, we apply the same fit, and after the reverse transformation we add the stored residual onto the track position, to obtain the original cluster coordinates in pad, row, and time.
Thus, the decompression is exact.

The prototype featuring all these improvements achieves a compression factor of~$9.1$ for Pb-Pb data from 2015 and~$9.0$ for pp data.
We mention that the only non-lossless compression steps are the cluster finding itself, and the data format conversion to fixed-point integer including the truncation for the cluster charge and width.
These formats are tuned to maintain the detector resolution.
All additional steps are lossless.

\looseness=-1
The number 9.1 still misses roughly a factor 2 to the required total compression of 20.
Therefore, the online reconstruction will identify low momentum tracks below 50\,MeV, additional legs of looping tracks, and track segments with high inclination angle, which are not used for physics.
Since roughly 50\,\% of the clusters belong to these cases, removing such clusters will improve the compression by the missing factor of 2.

\section{Online Calibration}

\looseness=-1
Online calibration is a special case in the current HLT processing.
While normal reconstruction is event synchronous, calibration tasks need many events to generate the calibration output, and it is impossible to cache the incoming data long enough.
Hence, the HLT employs an asynchronous scheme for online calibration, where all computing nodes perform the calibration task for a subselection of events they reconstruct~\cite{chep15}.
After enough events have been processed, all computing nodes send their calibration results to a common merger node, which merges the data and creates the final calibration objects.
These objects, like for the TPC transformation, are then redistributed in the cluster.
In this way, e.\,g.~the TPC drift time calibration computed for few minutes of data taking is used for the following minutes, which is possible if the drift velocity is stable during that time.
In order to use as much as possible the offline calibration procedure and avoid code duplication, the HLT employs a wrapper that can execute offline analysis tasks in the HLT framework.
However, instead of the normal ROOT-based ESD (Event Summary Data), the HLT passes a flat structure with slightly reduced content, which is sufficient for the calibration but avoids the serialization and deserialization of ROOT objects~\cite{cdt}.
With this approach, the HLT performs online TPC drift time calibration, where the final spacial cluster position deviate by less than 0.5\,mm from the position obtained via offline drift time calibration~\cite{tns2016}.

\label{sec:scd_calib}

\begin{figure}[b!]
 \begin{minipage}[t]{0.495\textwidth}
  \centering
  \includegraphics[height=2in]{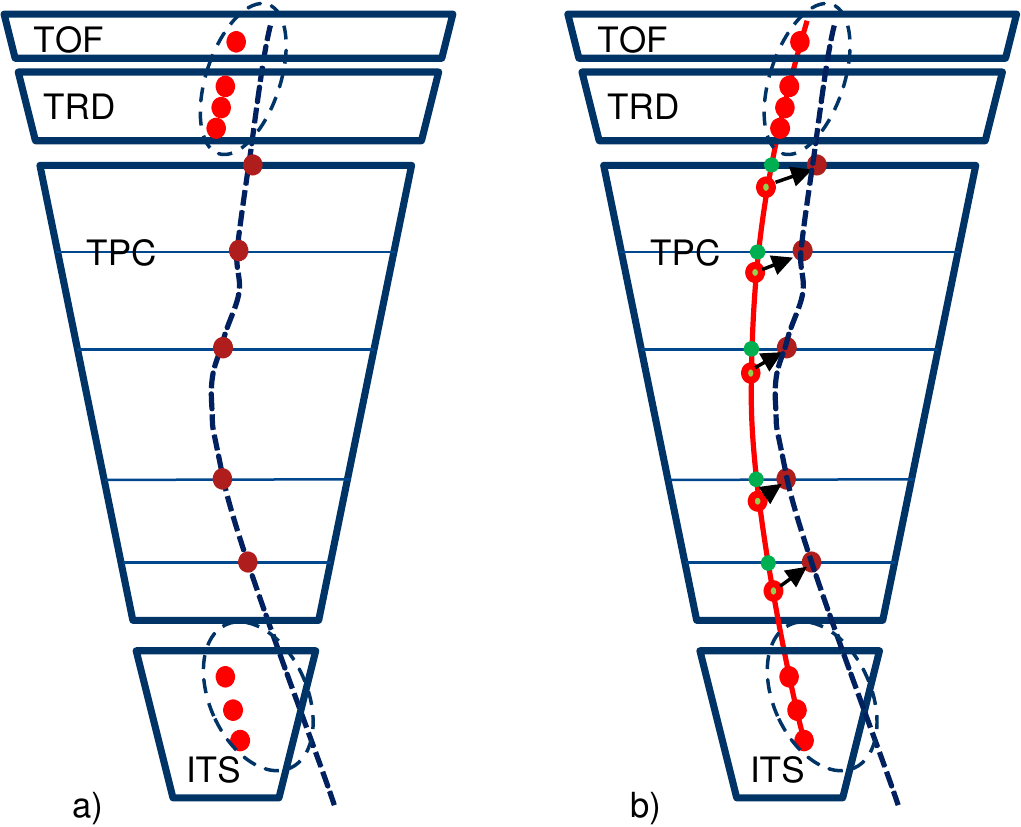}
  \caption{ Illustration of SCD calibration procedure }
  \label{fig:dist}
 \end{minipage}
 \begin{minipage}[t]{0.495\textwidth}
  \centering
  \includegraphics[height=2in]{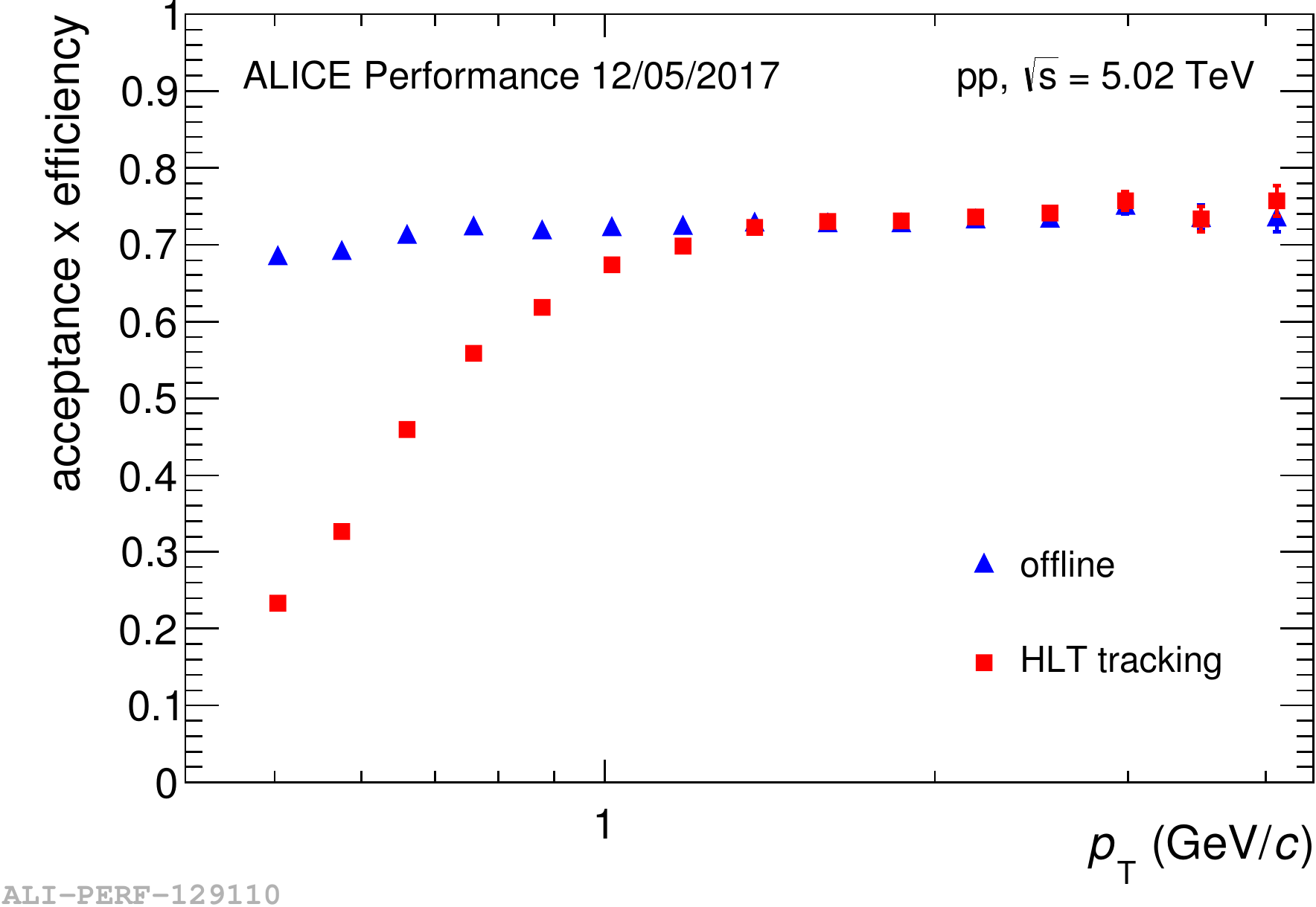}
  \caption{ HLT TRD Tracklet matching efficiency }
  \label{fig:trd}
 \end{minipage}
\end{figure}

\begin{figure}[htb]
\centering
\includegraphics[height=1.5in]{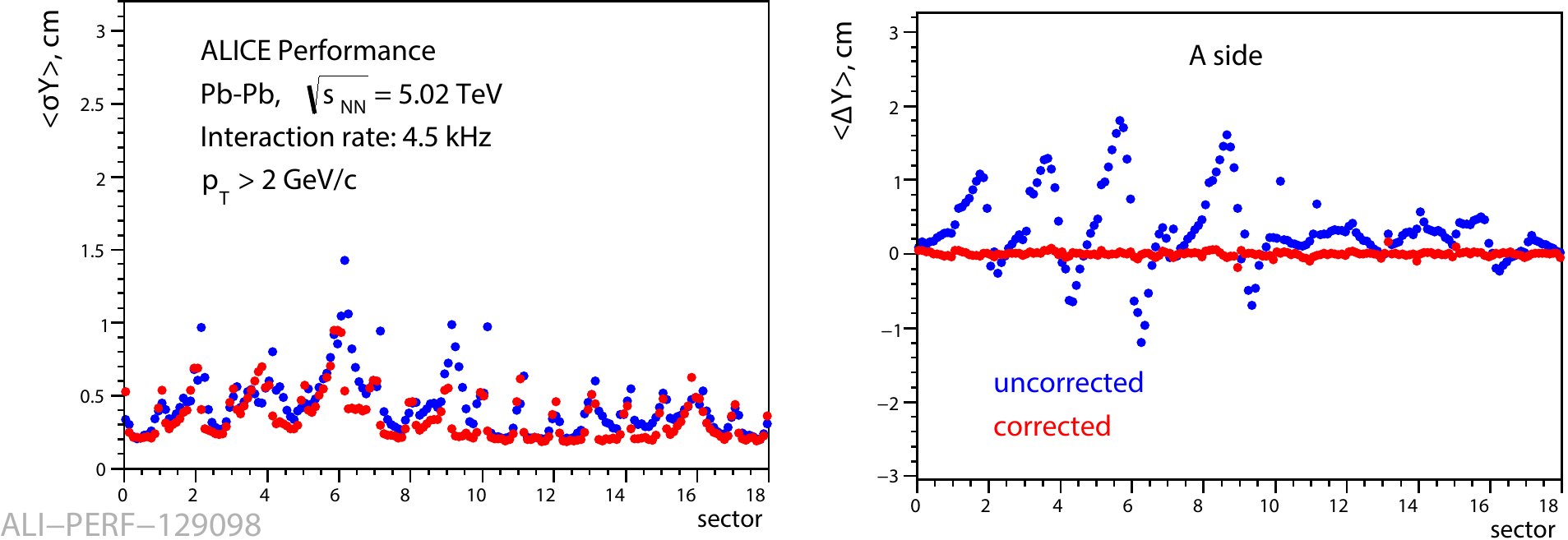}
\caption{ Effect of TPC SCD distortions correction on the DCA to the primary vertex using TPC-only track information (TPC A side only) }
\label{fig:tpc_corr}
\end{figure}

\looseness=-1
As next step, we will extend this procedure to the TPC SCD calibration.
Space charge of ions in the TPC distorts the electrons while they drift from the ionization point to the end plate.
These distortions scale with occupancy and have local effects.
The offline TPC reconstruction employs the following calibration procedure:
\begin{compactitem}
 \item The tracks in the TPC are reconstructed with relaxed tolerances and then matched to hits in the inner ITS detector and outer TRD and TOF detectors as illustrated in Fig.~\ref{fig:dist}a.
 \item The track is refitted using only ITS, TRD, and TOF information, and the residuals of the TPC hits to the refitted track are stored (Fig.~\ref{fig:dist}b).
 \item Having reconstructed sufficiently many events, a map of distortion corrections is obtained which in average corrects the TPC hits to lie on the ITS-TRD-TOF track.
 \item To have enough statistics, this procedure is done in 40 minute intervals.
 The TPC volume is voxelized and the correction is interpolated using Chebyshev polynomials.
\end{compactitem}
Fig.~\ref{fig:tpc_corr} shows how the SCD calibration corrects the bias of the DCA to the primary vertex.
There is still a degradation of the resolution in areas with large distortions because the 40 minute interval is too long to account for short fluctuations.
We plan to overcome this deficit in Run 3 by integrating the currents from the TPC read out, which will allow us to compute the local space charge and scale the corrections accordingly.
This approach cannot be implemented for Run 2 because the majority of the charge information is missing in the triggered read out.

Running this procedure online requires real-time tracking for the ITS, TRD, and TOF detectors.
While the ALICE HLT already performs ITS tracking, TRD tracking is currently under development as presented in the next section.

\section{Tracking with Continuous Read Out}

\looseness=-1
During run 3, the high data rate will prevent a full read out of all TRD hits.
Therefore, the online tracking that is developed will be based on local TRD online-tracklets found by the TRD read out electronics.
The current approach extrapolates TPC tracks through the TRD volume and attaches the closest tracklet.
For high multiplicity environments, this algorithm is currently refined to support a multiple track hypothesis approach.
Fig.~\ref{fig:trd} shows the tracking efficiency of online and offline TRD reconstruction.
The online version shows comparable matching efficiency for $P_{\text{T}} > 1$\,GeV.
There is a decrease for the online version at low momentum, but this is no deficiency of the tracking itself.
The read out hardware does not find low momentum online-tracklets because it was designed for high momentum triggers.
There is an ongoing effort to improve the TRD online-trackleting and decrease the threshold to~$0.6$\,GeV, which is needed for the SCD calibration.
The offline tracking is not affected because it produces offline tracklets from TRD hits, but these will not be available in Run 3.

\looseness=-1
The continuous read out poses additional challenges for the TPC reconstruction.
We will process full time frames instead of events with the GPU tracking, which solves the problems that events are too small to load the GPU efficiently.
Since the tracking time depends linearly on the input data size, it should not cause an overhead.
The limit is in fact the GPU memory, which will be sufficient to hold a full timeframe or a large part of it.
In the latter case, it is still possible to slice the timeframe in parts as we do with the TPC sectors today.

On top, the tracking needs to find and treat multiple vertices in a drift time.
The most prominent problem is that it is no longer possible to convert the time to a spacial coordinate before the tracking, because that conversion needs the time of the vertex, which is only known after reconstruction.
Therefore, the tracking will run in time coordinate first to estimate the vertex and only then convert to spacial coordinates, taking into account SCD.
Moreover, we plan to move additional reconstruction steps onto the GPU, starting with the TPC track fit, and later ITS and TRD tracking and d$E$/d$x$ calculation.

\section{Conclusions}

\looseness=-1
ALICE will take 50\,kHz of minimum bias Pb-Pb data during Run 3 and is updating its online and offline computing systems accordingly.
The new online farm will perform online reconstruction and calibration, and based on that, data compression by a factor of 20.
The new features are tested as far as possible in the current HLT online processing farm.
A prototype already reaches a compression factor of~9.1, and the missing factor of 2 will be obtained by rejecting TPC hits not used for physics.
Online calibration has been demonstrated for the TPC drift velocity and is currently extended to the space charge distortion calibration.
This requires online TRD tracking, which is currently developed.
Its prototype shows good matching efficiency but suffers from the lack of low-$P_{\text{T}}$ online TRD tracklets.
The TPC tracking, the biggest contributor to computing time, will be derived from the current HLT version based on the Cellular Automaton and Kalman filter.
Today, the HLT farm can track~$40,000,000$ tracks per second with equal efficiency as offline.
Some shortcomings in the fit degrading the resolution are currently addressed.
We will perform TPC tracking of full time frames on GPUs to speed up the processing, and we plan to offload additional reconstruction steps on GPUs as well.

\end{document}